\documentclass[prb,twocolumn,aps,showpacs,fixfloats]{revtex4}
\usepackage{graphicx}
\usepackage{bm}
\usepackage{amsmath,amssymb}
\usepackage{subfigure}
\usepackage{float}
\usepackage{latexsym}
\usepackage{color}
\usepackage{enumerate}
\usepackage{pdfpages}
\usepackage{tikz}
\usepackage{hyperref}
\usepackage{graphicx}
\usepackage{bm}
\usepackage{amssymb} 
\usepackage{amsmath}
\usepackage{subfigure}

\usepackage{relsize}

\begin{document}

\title{Effect of the resonant ac-drive on the spin-dependent recombination of  polaron pairs:  Relation to organic magnetoresistance }

\author{M. E. Raikh}
\affiliation{Department of Physics and
Astronomy, University of Utah, Salt Lake City, UT 84112}

\begin{abstract}
The origin of magnetoresistance is bipolar organic materials is the influence of magnetic field on the
dynamics of recombination within localized 
electron-hole pairs. Recombination from the $S$ 
spin-state  of the pair in preceded by the beatings between the states $S$ and $T_0$. Period of the beating is set by the the random hyperfine field. For the case when  recombination time from $S$ is shorter than the period, we demonstrate that a {\em weak} resonant ac drive, which couples $T_0$ to $T_+$
and $T_{-}$ affects dramatically the recombination
dynamics and, thus, the current A distinctive characteristics of the effect is that the current versus the drive amplitude exhibits a {\em maximum}. 
\end{abstract}

\maketitle

\section{Introduction}
 In the pioneering paper Ref. \onlinecite{Frankevich}
it was first proposed that sensitivity of the resistance of certain organic materials to a weak magnetic field is due to spin-dependent recombination within polaron pairs. Roughly speaking, while formation of a pair in each of the four possible spin states $S$, 
$T_0$, $T_+$, and $T_-$, occurs with equal probabilities, they recombine predominantly from $S$. As a result, the $S-T$ dynamics ($S-T$ beatings) affects the net recombination rate, which, in turn, determines the current. On the other hand, the $S-T$ dynamics (see the classical paper Ref. \onlinecite{classical}) is governed by the ratio of the applied magnetic field to the hyperfine field.
If both pair-partners have the same charge, recombination should be replaced by the bipolaron formation.\cite{VardenyUltrasmall}
In the later studies (see e.g. Refs. \onlinecite{OMAR1,OMAR2,OMAR3,OMAR4,Roundy1,Roundy2}) the scenario of Ref. \onlinecite{Frankevich} was confirmed experimentally. A conclusive experimental
evidence 
that hyperfine magnetic fields play a crucial role in
magnetic field response of organic semiconductors
was reported in Ref.~\onlinecite{5} on the basis of
comparison  the data on hydrogenated (with
higher hyperfine fields) and deuterated (with lower
hyperfine fields) samples.

It is clear that subjecting a pair to a resonant ac drive can flip the spin of one of the pair-partners and, thus, affect the $S-T$ beatings. 
This idea was proposed in Refs.~\onlinecite{Robust,Roundy}
and confirmed experimentally in Refs. 
\onlinecite{Experiment},\onlinecite{Experiment1},\onlinecite{Boehme+Lupton}.
  A peculiar feature of the analytical theory Ref. \onlinecite{Roundy} is that it predicted a non-monotonic dependence of magnetoresistance on the drive amplitude, $B_1$. This behavior reflects a delicate balance between two processes. Firstly, without drive, the only triplet state coupled to $S$ is $T_0$. In other words, it cannot recombine from $T_+$ or $T_{-}$, which we call the trapping configurations. Weak drive couples $T_+$ and $T_{-}$ to $T_0$, thus, effectively coupling them to $S$ and allowing recombination from these states.  This leads to the enhancement of current. On the other hand, strong drive leads to the formation of a new mode of spin dynamics\cite{Roundy}, namely, $T_+ -T_-$. This mode is orthogonal to $T_0$, and, thus, it is decoupled from $S$. Possibility to be trapped in $T_+ -T_-$
  leads to the slowing of recombination and, correspondingly, to the decrease of current upon increasing of drive.

The theory of Ref. \onlinecite{Roundy} was based on the assumption that the $S-T$  asymmetry of the pair is strong. Typical value of this asymmetry is $\gamma b_0$, where $\gamma$ is the gyromagnetic ratio, while $b_0$ is the magnitude of the hyperfine field. The criterion of a strong asymmetry reads $\gamma b_0\tau \gg 1$, where $\tau$ is the recombination time from $S$. Physically, this criterion implies that the pair undergoes many, $\sim \gamma b_0\tau$, \hspace{3mm} $S-T_0$ beatings before it recombines. In the opposite limit, $\gamma b_0\tau \ll 1$, there are no $S-T_0$ beatings. Instead, the pair recombines almost instantly after it finds itself in $S$. In this limit it is not clear whether the ac drive can affect the current. It is this limit that is studied in the present paper. We employ the same maximally simplified transport model as in Refs. \onlinecite{Roundy1,Roundy2}.
\begin{figure}
\label{1}
\includegraphics[scale=0.38]{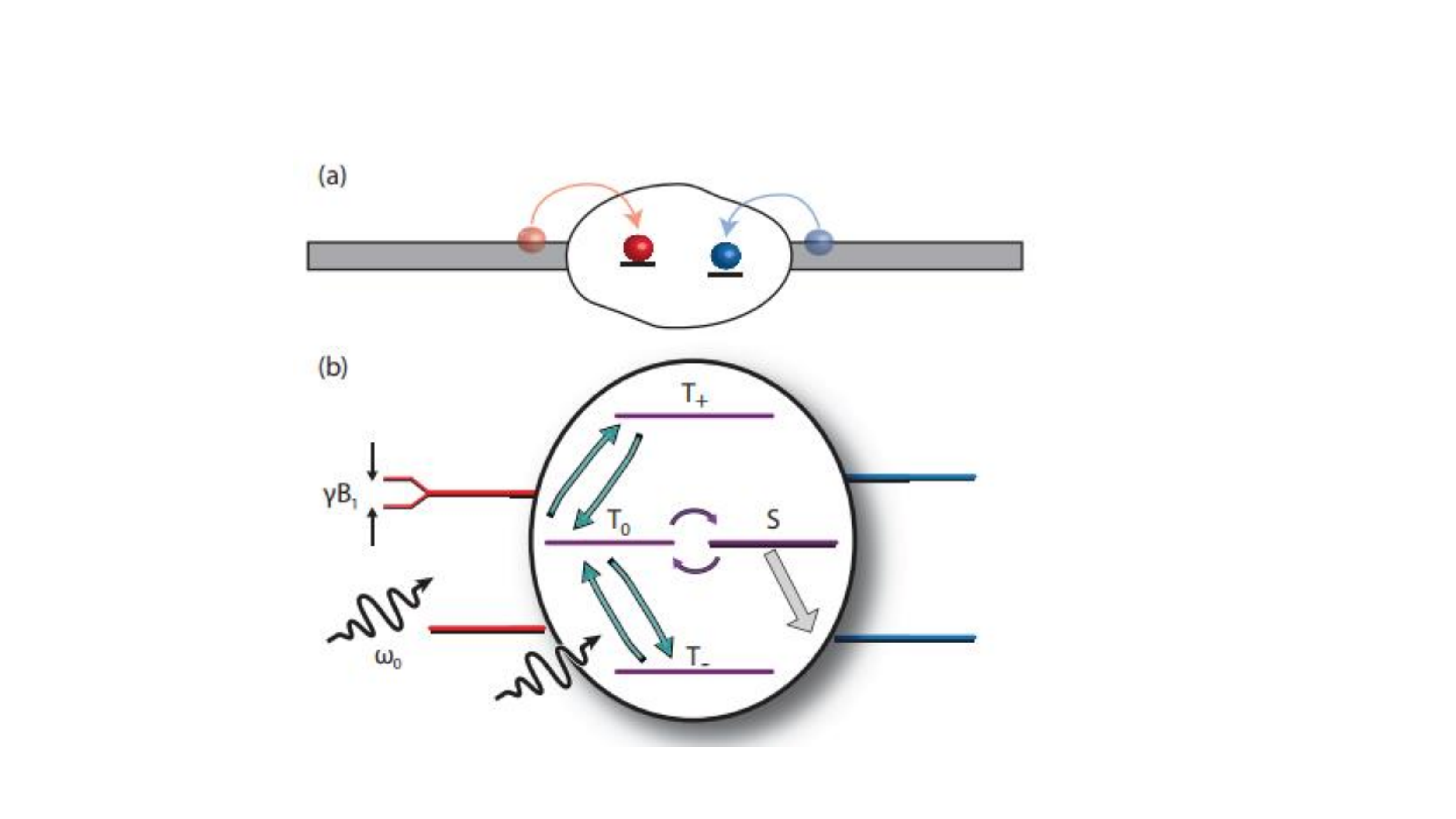}
\caption{(Color online)  (a) Illustration of the transport model. Current passage
 through a bipolar device involves recombination of the localized electron (red) and localized hole (blue), which are assembled on the neighboring sites; 
(b) Spin dynamics of a resonantly driven pair preceding recombination from $S$. The dynamics involves $S-T_0$ beatings and the Rabi oscillations between
$T_0$ and two other triplet components.   
}
\end{figure}

\section{Transport model}

For concreteness we assume that the transport is due to recombination of electron-hole pairs. As illustrated in 
Fig.~1, electron and hole are assembled at neighboring sites and start precessing in the respective magnetic fields, ${\bm B}+{\bm b_e}$ and ${\bm B}+{\bm b_h}$, where ${\bm b_e}$ and ${\bm b_h}$ are the electron and hole hyperfine fields. All four initial spin configurations of a pair have the same probability. After some time, the pair either recombines or gets disassembled depending on which process takes less time.  The key simplifying assumption adopted in Refs. 
 \onlinecite{Roundy1,Roundy2}      is that assembly and disassembly of a pair takes the time, $\tau_D$, much longer than $\tau$. Under this assumption, the passage of current proceeds in cycles. The magnitude of current is related to 
 $\langle t \rangle$  which is the average cycle duration, see e.g. Ref.
 \onlinecite{von}, as
\begin{equation}
\label{current}
I=\frac{1}{\langle t \rangle}.
\end{equation}
Obviously, the ac drive, $B_1$, affects only the precession stage of the current passage.  The effect of drive is most pronounced when $|{\bm b}_e|, |{\bm b}_h| <<B$. Then the resonant frequency is related to $B$ as, $\omega = \gamma B$. The above condition allows to neglect the transverse components of hyperfine fields and to employ the rotating wave approximation.

\section{Spin dynamics with fast recombination}
We start from the standard Hamiltonian of an electron-hole pair driven with frequency $\omega = \gamma B$
\begin{equation}
\label{Hamiltonian}
\nonumber
\!{\hat H}=\omega_e{\hat S}_e^z+\omega_h{\hat S}_h^z+2\Omega_R\left({\hat S}_e^x+{\hat S}_h^x\right)\cos\omega t,
\end{equation}
where $\Omega_R=\gamma B_1$ is the Rabi frequency, ${\hat S}_e$ and 
${\hat S}_h$ are the spin operators of electron and hole, respectively. The frequencies $\omega_e$ and $\omega_h$ are expressed via the $z$-projections of the hyperfine fields $b_e^z$, $b_h^z$ as

\begin{equation}
\label{definition}
\omega_e=\omega +\gamma b_e^z, \hspace{4mm}
\omega_h=\omega+\gamma b_h^z. 
\end{equation}
Denote with $A_{T_+}$, $A_{T_{-}}$, $A_{T_0}$, and $A_S$ the amplitudes of four spin states of the pair.  With account of recombination from $S$, the amplitudes of the states satisfy the system of equations
\begin{align}
\label{system}
&i\frac{\partial A_{T_+}}{\partial t}=\left(\omega+\delta \right) A_{T_+}
+ \frac{\Omega_R}{2^{1/2}}A_{T_0} \cos\omega t,\\
&i\frac{\partial A_{T_{-}}}{\partial t}=-\left(\omega+\delta \right) A_{T_{-}}
+ \frac{\Omega_R}{2^{1/2}}A_{T_0} \cos\omega t,\\
&i\frac{\partial A_{T_0}}{\partial t}=\delta_0A_S+\frac{\Omega_R}{2^{1/2}}
(A_{T_+}+A_{T_-}   )\cos\omega t,\\
&i\Bigl(\frac{\partial A_S}{\partial t} +\frac{A_S}{\tau}\Bigr) =\delta_0A_{T_0},
\end{align}
where $\delta_0$ and $\delta$ are defined as 
\begin{equation}
\label{delta}
\delta_0 =\frac{1}{2}\left(\omega_e - \omega_h\right) , \hspace{3mm} 
\delta = \frac{1}{2}\left(\omega_e + \omega_h -2\omega\right).
\end{equation}
Within the rotating wave approximation 
$\cos\omega t$ in the first equation of the system is replaced by 
$\frac{1}{2}e^{-i\omega t}$, while in the second equation it is replaced by $\frac{1}{2}e^{i\omega t}$. Switching to “rotated” system is achieved by introducing new variables
\begin{align}
\label{variables}
& A_{T_{-}} =a_{T_{-}}e^{-i(\chi-\omega)t}, \hspace{3mm}
A_{T_{+}} =a_{T_{+}}e^{-i(\chi+\omega)t}, \\
& A_{T_{0}} =a_{T_{0}}e^{-i\chi t}, \hspace{3mm} A_S=a_Se^{-i\chi t}.
\end{align}
Then the system reduces to the following four algebraic equations
\begin{align}
\label{a}
& (\chi-\delta)a_{T_{-}}=\frac{\Omega_R}{2^{1/2}}a_{T_0}, \hspace{2mm}
(\chi+\delta)a_{T_{+}}=\frac{\Omega_R}{2^{1/2}}a_{T_0},\\ \label{a1}
&\chi a_{T_0} =-\delta_0a_S+\frac{\Omega_R}{2^{1/2}}(a_{T_{-}}+a_{T_{+}}), \nonumber\\ 
&\left(\chi +\frac{i}{\tau}\right)a_S=-\delta_0a_{T_0},
\end{align}
As discussed above, in the absence of drive, spin dynamics of the system reduces to the $S-T$ beating with a frequency $\delta_0$, which is a measure of the hyperfine-induced spin asymmetry of the pair. 
Indeed, upon setting $\Omega_R=0$ in the system Eqs. (\ref{a}), (\ref{a1}) it yields two frequencies, $\pm \delta_0$.
Subjecting the system to the ac drive, involves the states $T_+$ and $T_{-}$ into the spin dynamics leading to the formation of two more modes. The frequencies of all four modes satisfy the following algebraic equation, which follows from Eqs. (\ref{a}),\hspace{1mm}(\ref{a1})
\begin{equation}
\label{chi}
\chi\Bigl(\chi+\frac{i}{\tau}\Bigr)\Bigl(\chi^2-\delta^2-\Omega_R^2\Bigr)=\delta_0^2\Bigl(\chi^2-\delta^2\Bigr). 
\end{equation}
Unlike Ref. \onlinecite{Roundy}, we will analyze the eigenvalues of the system Eqs. (\ref{a}),
\hspace{1mm} (\ref{a1}) in the limit of fast recombination  $\tau^{-1} \gg \delta_0,\delta,\Omega_R$. One
consequence of the fast recombination is
that Eq. (\ref{chi}) has a solution 
$\chi=-\frac{i}{\tau}$ corresponding to
the decaying state $S$. Then the other three solutions
for which $|\chi|\ll \tau^{-1}$ satisfy the cubic equation
which emerges upon neglecting $\chi$ in the second bracket
\begin{equation}
\label{cubic}    
\chi^3-\Bigl(\delta^2+\Omega_R^2\Bigr)\chi=-i\delta_0^2\tau\Bigl(\chi^2-\delta^2\Bigr).
\end{equation}
Another consequence of $\tau^{-1}$ being bigger than other parameters in Eq. (\ref{chi}) is that one of three brackets in the left-hand side of Eq. (\ref{cubic}) is  small.  Thus, in the limit $\tau \rightarrow 0$ the three eigenvalues are
\begin{equation}
\label{three}
\tilde{\chi}_{T_0}=0, \hspace{2mm}
\tilde{\chi}_\pm = \pm(\delta^2+\Omega_R^2)^{1/2}. 
\end{equation}

\begin{figure}
\label{2}
\includegraphics[scale=0.42]{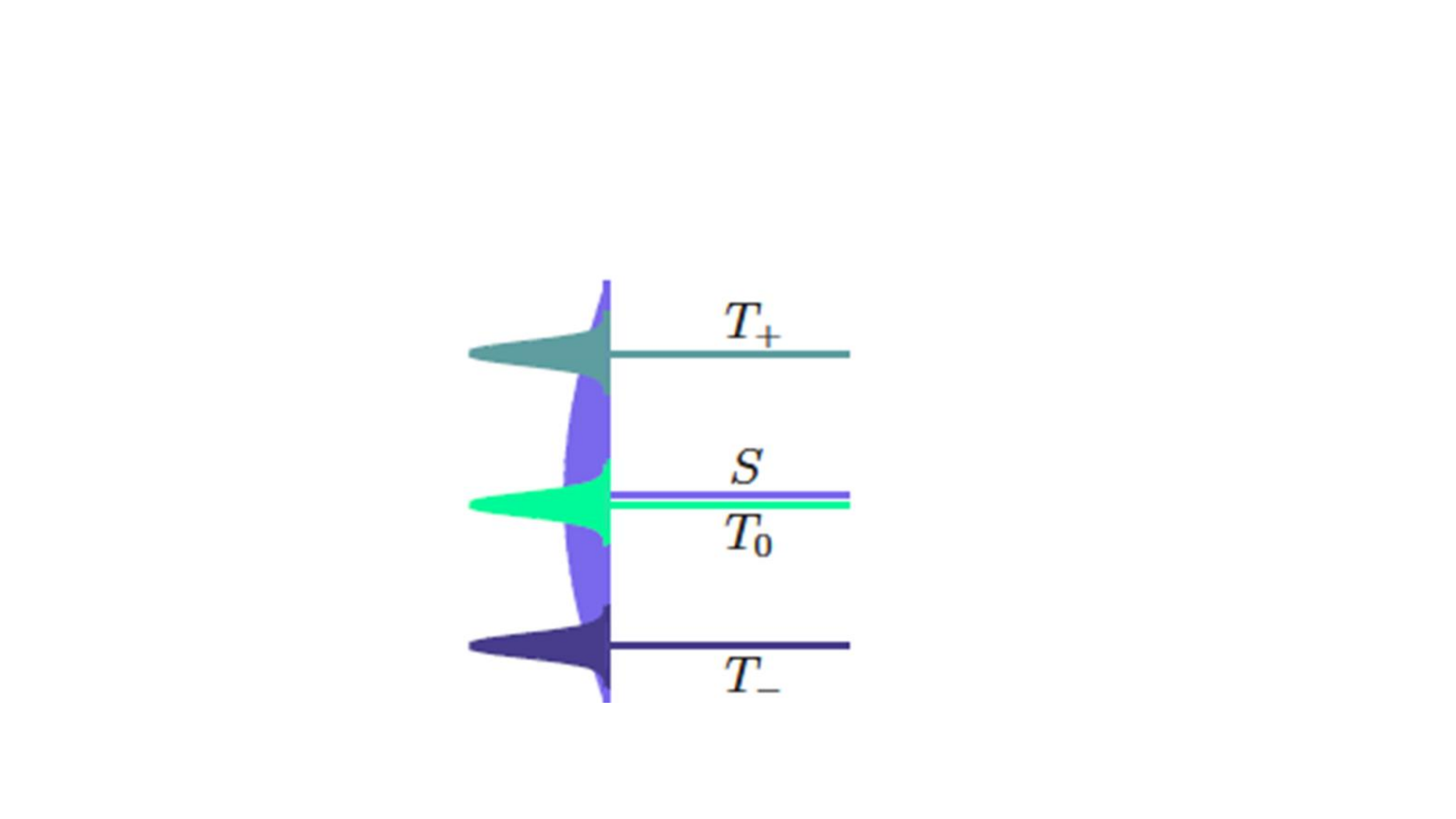}
\caption{(Color online) Schematic illustration of the
eigenvalues of Eq. (\ref{chi}) in the regime of fast recombination.
The width, $\tau^{-1}$, of the eigenstate $S$ is much bigger than
the widths of $T_0$, $T_+$, $T_{-}$ eigenstates.}
\end{figure}

Naturally, the zero-order eigenvalues Eq. (\ref{three}) do not capture the decay of $T_0$ and $T_{\pm}$ eigenmodes. This is because we neglected the right-hand side in Eq. (\ref{cubic}) containing the recombination time.
 To capture the decay of eigenmodes, we substitute 
the zero-order $\tilde{\chi}$ values into the right-hand side, set 
 $\chi=\tilde{\chi} +\delta \chi$ in the left-hand side, and expand the left-hand
 side with respect to $\delta\chi$. This yields
 \begin{equation}
 \label{correction}
 \delta\chi=-i\delta_0^2\tau
\left( \frac{\tilde{\chi}^2-\delta^2}
 { 3\tilde{\chi}^2-\delta^2-\Omega_R^2           }\right)
 \end{equation}
 Using Eq. (\ref{three}), we find the following imaginary parts of slow-decaying
modes
\begin{equation}
 \label{imaginary}
 \delta\chi_{T_0}=-i\delta_0^2\tau
\left(\frac{\delta^2}{\delta^2+\Omega_R^2}\right),
\end{equation}
\begin{equation}
\label{imaginary1}
 \delta\chi_{\pm} = -i\delta_0^2\tau
\left(\frac{\Omega_R^2}{2\delta^2+2\Omega_R^2}\right).
\end{equation}
By virtue of the small parameter $\delta_0\tau\ll 1$ the
widths of the triplet eigenmodes is much smaller than $\tau^{-1}$, the width of the singlet eigenmode, as illustrated
in Fig. 2.
From Eqs. (\ref{imaginary}), (\ref{imaginary1}) we
conclude that
\begin{equation}
\label{conclude}
\delta\chi_{T_0} +\delta\chi_{+}+\delta\chi_{-}=-i\delta_0^2\tau, 
\end{equation}
i.e. the sum of the decay rates of
weakly decaying modes is {\em  independent  of drive.}
The main point though is that the current is determined
by the average {\em trapping time}, while  contribution to the trapping time from each mode is proportional to the 
{\em inverse} decay rate. Using Eqs.
(\ref{imaginary}), (\ref{imaginary1}) we
can express the average duration of a 
cycle as follows
\begin{equation}
\label{main}
\langle t \rangle-\tau_D=\sum_i\frac{1}{4\text{Im}~\chi_i}=
\frac{\tau}{4}+\frac{\delta^2+\Omega_R^2}{4\delta_0^2\tau}
\left[\frac{4}{\Omega_R^2}+\frac{1}{\delta^2}\right].
\end{equation}
The right-hand side of Eq. (\ref{main})
can be presented in the form
\begin{equation}
\label{as}
\frac{\tau}{4} +\frac{1}{4\delta_0^2\tau}\left[      \frac{4\delta^2}{\Omega_R^2}+\frac{\Omega_R^2}{\delta^2} +5\right].
\end{equation} 
This form suggests that the period 
of the cycle and, thus, the inverse current as a function
of drive has a minimum at $\Omega_R=2^{1/2}\delta$. Recall,
that the parameter $\delta$, defined by Eq.~ (\ref{delta}), has the meaning of the combined detuning of the pair.

 In principle, the term $\tau$ in Eq. (\ref{main}) also contains a $\Omega_R^2$ correction. To calculate this correction we recast Eq. (\ref{chi}) in the form
\begin{equation}
\label{recast}
\chi+\frac{i}{\tau}=\frac{\delta_0^2(\chi^2-\delta^2)}
{\chi(\chi^2-\delta^2-\Omega_R^2)}.
\end{equation}
Substituting the zero-order result, $\chi=-\frac{i}{\tau}$,
into the right-hand side, we get a corrected expression for
the imaginary eigenvalue
\begin{equation}
\label{correction}
 \chi_S=-\frac{i}{\tau}\left[1-
\frac{\delta_0^2\tau^2(1+\delta^2\tau^2)}{1+(\delta^2+\Omega_R^2)\tau^2          }\right].   
\end{equation}
It follows from Eq. (\ref{correction})
that the magnitude of $\Omega_R$-dependent correction to the decay time
of the $S$-mode is $\sim \delta_0^2\tau^3$, which is much smaller than the remaining terms in Eq. (\ref{as}) by virtue of the small parameter $\delta_0\tau \ll 1.$
 
\section{Incorporating exchange}
Next question we address is how the exchange
interaction between electron and hole affects
the current through the resonantly driven pair.
For a standard exchange Hamiltonian 
\begin{equation}
\label{exchange}
  {\hat H}_{ex}=-J{\bf {\hat S}}_e\cdot{\bf {\hat S}}_h
\end{equation}
the system Eq. (\ref{a}) is modified as follows.
In equations for $T_0$, $T_\pm$ the eigenvalue
$\chi$ is replaced by 
$\bigl(\chi +\frac{J}{2}\bigr)$, while in equation for $S$ it is replaced by $\bigl(\chi -\frac{J}{2}\bigr)$.
Then in the limit of fast recombination 
Eq. (\ref{cubic}) takes the form
\begin{align}
\label{cubic1}    
&\left(\chi+\frac{J}{2}\right)^3-\left(\delta^2+\Omega_R^2\right)\left(\chi+\frac{J}{2}\right)\nonumber\\
& =-i\delta_0^2\tau\left[\left(\chi+\frac{J}{2}\right)^2-\delta^2\right].
\end{align}
From the form of Eq. (\ref{cubic1}) we conclude
that
incorporating exchange shifts the eigenvalues by $J$ but leaves their imaginary parts, 
responsible for the current, unchanged.

\section{Averaging over the hyperfine fields}

Averaging over the hyperfine fields acting on electron and hole
reduces to the averaging of Eq. (\ref{as}) over parameters $\delta$
and $\delta_0$, which are statistically independent and
Gauss-distributed. Prior to performing the averaging,
we recast the expressions Eqs. (\ref{current}) and (\ref{as}) in the form

\begin{equation}
\label{Average0}
I(\Omega_R)=\frac{\Omega_R^2\tau}{a\Omega_R^4+b\Omega_R^2+c},
\end{equation}
where the parameters $a$, $b$, and $c$ are expressed via the following averages
\begin{align}
\label{abc}
 & a=\frac{1}{4} \Big{\langle} \frac{1}{\delta^2\delta_0^2}\Big{\rangle},\\
 & b=\tau \biggl(\tau_D+\frac{\tau}{4}\biggr)+
 \frac{5}{4}\Big{\langle} \frac{1}{\delta_0^2}  \Big{ \rangle},\\
 & c=\Big{\langle} \frac{\delta^2}{\delta_0^2}   \Big{  \rangle}.
\end{align}
Note, that all three parameters {\em diverge} in the limit of small $\delta$, $\delta_0$. These divergences should be cut at
$\delta, \delta_0 \sim \tau_D^{-1}$, i.e. at inverse assembly time
of the pair. The portion of such small
$\delta, \delta_0$ is $(\tau_D\gamma b_0)^{-1}$, where $\gamma b_0$ is the r.m.s. detuning of the pair partners. With this in mind, we get the following expressions for the parameters $a$,
$b$, and $c$
\begin{equation}
\label{ABC}
a=\frac{1}{4}\left(\frac{\tau_D}{\gamma b_0}\right)^2,
~~b=\frac{5}{4}\left(\frac{\tau_D}{\gamma b_0}\right),~~
c=\gamma b_0\tau_D.
 \end{equation}
 It is instructive to present the final result 
 Eq. (\ref{Average0})
in terms of dimensionless Rabi frequency defined as
\begin{equation}
 \label{Dimensionless}
{\tilde \Omega_R}=\Omega_R\frac{\tau_D^{1/4}}{(\gamma b_0)^{3/4}}.
\end{equation}
Then Eq. (\ref{Average0}) takes the form
\begin{equation}
 \label{final}
 I(\Omega_R)=\frac{\tau(\gamma b_0)^{1/2}}{\tau_D^{3/2}
\left[\frac{1}{4}{\tilde \Omega_R}^2 +{\tilde \Omega_R}^{-2}      +\frac{5}{4}(\gamma b_0 \tau_D)^{-1/2}            \right]}.
\end{equation}
The fact that within the adopted transport model the assembly time, $\tau_D$, exceeds
the precession time, $\tau_D\gg (\gamma b_0)^{-1}$. One consequence of this assumption
is the third term in the denominator of Eq. (\ref{final}) is smaller than one, i.e. the
the minimum is deep. Another nontrivial consequence of a long assembly time is that the position $\tilde \Omega_R \sim 1$
of maximum of $I(\Omega_R)$ corresponds to the Rabi frequency
 {\em parametrically smaller} than the broadening, $\gamma b_0$, 
 due to hyperfine fields.
 
 In Eq. (\ref{definition})
we assumed that the driving frequency, $\omega$, exactly matches
the Zeeman splitting, $\gamma B$. In fact, in 
the experiment of Ref. \onlinecite{Robust} the sensitivity  of current 
to the ac drive was observed in the sizable interval of $B$.
 The result of averaging over $\delta$ depends on the detuning 
\begin{equation}
\label{Delta}
\Delta=\omega-\gamma B.
\end{equation}
In the case when the detuning is big, 
$\Delta \gg \gamma b_0$, one has to replace
$\delta$ by $\Delta$ on Eq. (\ref{abc}) and to
perform averaging over $\delta$. The result reads
\begin{equation}
 \label{FINAL}
 I(\Omega_R)=\frac{4\gamma b_0\tau}
 {\tau_D \left[\frac{\Omega_R^2}{\Delta^2} + \frac{4\Delta^2}{\Omega_R^2} +5   \right]}.
\end{equation}

\section{Discussion}
({\bf 1}). Our main result is Eq. (\ref{as}) for the  time, $\langle t \rangle$,  as a function
of the drive amplitude, $\Omega_R=\gamma B_1$.
The meaning of $\langle t \rangle$ is 
the time it takes for a spin  pair to assemble and either recombine or disassemble. Our main finding
is that, in the regime of fast recombination,
$\langle t \rangle$,
 exhibits 
a minimum. Thus, the current $I=\langle t \rangle^{-1}$ versus $\Omega_R$ exhibits
a maximum in agreement with slow-recombination theory.\cite{Roundy} It should be emphasized that the faster
is recombination, the {\em smaller} is the current. This is
because {\em fast} recombination from $S$ ensures more efficient
trapping in each of the triplet states.

({\bf 2}). We have incorporated decay into the system
Eq.~(\ref{a}) for the {\em amplitudes} of four states of the spin-pair. Rigorous approach implies incorporation of the decay into  the  stochastic Liouville equation for the spin density matrix. With four spin states involved, the size of the density matrix is
$16\times 16$. However, the eigenvalues of this matrix can be cast in the form $\chi_i-\chi_j^{*}$, where $\chi_i$ are the roots of the fourth-order algebraic equation Eq. (\ref{chi}). Equivalence of the description adopted in the present paper and the
density matrix-based description was demonstrated in
Refs. \onlinecite{Roundy1,Shinar}. Roughly speaking, 
our description applies when the eigenvectors of the
$4\times 4$ system for the amplitudes are approximately
orthogonal to each other. In our case, approximate orthogonality of eigenvectors
is insured by the fact that one eigenvector is mostly of $S$-type with small
(of the order of $\delta_0\tau$)
admixture of $T$-states, while other eigenstates are mostly $T$ with small admixture
of $S$

({\bf 3}). Consideration in the present paper pertains
to organic semiconductors since the spin-orbit coupling in carbon-based materials is weak.
The residual spin-orbit interaction renormalizes 
the difference, $\Delta g$, of the Land{\'e}
factors of the pair partners. In strong external
field the $S-T_0$ splitting caused  by $\Delta g$
exceeds the hyperfine-induced splitting, $\delta_0$, defined by Eq. (\ref{delta}). In 
Ref.~\onlinecite{deltaG} it was 
demonstrated experimentally that this 
$\Delta g$-mechanism is at work in the field $B\sim 8T$.

({\bf 4}). As it was mentioned in the Introduction, the 
 underlying mechanism of the sensitivity of current through the
 organic material to the magnetic field is the $S-T_0$ beating preceding
 the recombination from $S$. Change of the external field affects this beating. 
 Naturally, recombination causes the broadening
 of both $S$ and $T_0$ levels. Importantly, this broadening is 
 quite different in the limits of the slow, 
 $\gamma b_0\tau \gg 1$, and fast
 $\gamma b_0\tau \ll 1$  recombination. In the first limit both $S$ and $T_0$
 have equal widths, $\frac{1}{2\tau}$. In the second limit of the fast recombination 
 the width of $S$ 
 remains  $\frac{1}{2\tau}$, while the level $T_0$ narrows dramatically and 
 becomes $\gamma^2b_0^2\tau$, as it is illustrated in Fig.~2. In this regard, the key message of the present paper,
is that, with fast recombination, a narrow level $T_0$
can be engaged into the Rabi oscillations between $T_0$ and $T_+$, $T_{-}$
by a {\em{weak}} ac drive. Certain confirmation that this scenario is realistic comes from the fact that these Rabi oscillations can be "seen"
experimentally by means of the pulsed EDMR technique, see e.g.  Refs. \onlinecite{Boehme+Lips,Boehme+Lupton+Rabi,HighB_0,Saam,Glenn}.
Within this technique,
the change, $\delta\sigma$, in the conductivity of the sample is
measured upon application of a pulse of variable duration. Then the Rabi oscillations show up in the dependence of $\delta\sigma$  on the pulse duration.



\begin{thebibliography}{20}
\bibitem{Frankevich}
E. L. Frankevich, A. A. Lymarev, I. Sokolik, F. E. Karasz, S. Blmstengel, R. H. Baughman, and H. H. H{\"o}rhold, “Polaron-pair generation in poly(phenylene vinylenes),” 
Phys. Rev. B {\bf 46}, 9320 (1992).
\bibitem{classical}
K. Schulten and P. G. Wolynes, 
"Semiclassical description of electron spin motion in radicals including the effect of electron hopping,"
J. Chem. Phys. {\bf 68}, 3292 (1978).

\bibitem{VardenyUltrasmall}
Experimental evidence that the spin pairs can consist 
both of opposite-charge carriers (in bipolar devices) and
the like-charge carriers (in unipolar devices) can be found in
T. D. Nguyen, B. R. Gautam, E. Ehrenfreund, and Z. V. Vardeny,
"Magnetoconductance Response in Unipolar and Bipolar Organic Diodes at Ultrasmall Fields,"
Phys. Rev. Lett. {\bf 105}, 166804 (2010).
\bibitem{OMAR1}
{\"O}. Mermer, G. Veeraraghavan, T. L. Francis, Y. Sheng, D. T. Nguyen, M. Wohlgenannt, A. K{\"o}hler, M. K. Al-Suti, and M. S. Khan, “Large magnetoresistance in nonmagnetic $\pi$-conjugated semiconductor thin film devices,” Phys. Rev. B {\bf 72}, 205202 (2005). 
\bibitem{OMAR2}
S. P. Kersten, A. J. Schellekens, B. Koopmans, and P. A. Bobbert, "Magnetic-Field Dependence of the Electroluminescence of Organic Light-Emitting Diodes: A Competition between Exciton Formation and Spin Mixing,"
Phys. Rev. Lett. {\bf 106}, 197402 (2011). 
\bibitem{OMAR3}
W. Wagemans, A. J. Schellekens, M. Kemper, F. L. Bloom, P. A. Bobbert, and B. Koopmans, "Spin-Spin Interactions in Organic Magnetoresistance Probed by Angle-Dependent Measurements,"
Phys. Rev. Lett. {\bf 106}, 196802 (2011).

\bibitem{OMAR4} 
N. J. Harmon and M. E. Flatt{\'e},
"Spin-Flip Induced Magnetoresistance in Positionally Disordered Organic Solids,"
Phys. Rev. Lett. {\bf 108}, 186602 (2012);
"Semiclassical theory of magnetoresistance in positionally-disordered organic 
semiconductors,"
Phys. Rev. B {\bf 85}, 075204 (2012); 
"Effects of spin-spin interactions on magnetoresistance in disordered organic semiconductors," Phys. Rev. B {\bf 85}, 245213 (2012).
\bibitem{Roundy1}R. C. Roundy and M. E. Raikh, “Slow dynamics of spin pairs in a random hyperfine field: Role of inequivalence of electrons and holes in organic magnetoresistance,” Phys. Rev. B {\bf 87}, 195206 (2013).

\bibitem{Roundy2} R. C. Roundy, Z. V. Vardeny, and M. E. Raikh, “Organic magnetoresistance near saturation: Mesoscopic effects in small devices,” Phys. Rev. B {\bf 88}, 075207  (2013).

\bibitem{5} T. D. Nguyen, G. H. Markosian, F. Wang, L. Wojcik, X. G. Li, E. Ehrenfreund, and Z. V. Vardeny,
"Isotope effect in spin response of $\pi$-conjugated polymer films and devices,"
Nature Mater. {\bf 9}, 345 (2010).

\bibitem{Robust} 
W. J. Baker, K. Ambal, D. P. Waters, R. Baarda, H. Morishita, K. van Schooten, D. R. McCamey, J.M. Lupton, and C. Boehme,
"Robust absolute magnetometry with organic thin-film devices," Nat. Commun. {\bf 3}, 898 (2012).


\bibitem{Roundy} R. C. Roundy and M. E. Raikh, “Organic magnetoresistance under resonant ac drive,” Phys. Rev.  B {\bf 88}, 125206 (2013). 
\bibitem{Experiment}
D. P. Waters, G. Joshi, M. Kavand, M. E. Limes, H. Malissa, P. L. Burn, J. M. Lupton, and C. Boehme, “The spin-Dicke effect in OLED magnetoresistance,” Nat. Physics {\bf 11}, 910 (2015).
\bibitem{Experiment1}
S. Jamali, G. Joshi, H. Malissa, J. M. Lupton and C. Boehme,
"Monolithic OLED-Microwire Devices for Ultrastrong Magnetic
Resonant Excitation," Nano Lett. {\bf 17},  4648 (2017).

\bibitem{Boehme+Lupton}
T. H. Tennahewa, S. Hosseinzadeh, S. I. Atwood, H. Popli, H. Malissa, J. M. Lupton,
and  Boehme, 
"Spin relaxation dynamics of radical-pair processes at low magnetic fields,"
arXiv:2207.07086.

\bibitem{von} J. Koch, M. E. Raikh, and F. von Oppen,
"Full Counting Statistics of Strongly Non-Ohmic Transport through Single Molecules," 
Phys. Rev. Lett. {\bf 95}, 056801 (2005).







\bibitem{Shinar}
V. V. Mkhitaryan, D. Danilovi{\'c}, C. Hippola, M. E. Raikh, and J. Shinar,
"Comparative analysis of magnetic resonance in the polaron pair recombination and the triplet exciton-polaron quenching models,"
Phys. Rev. B {\bf 97}, 035402 (2018).

\bibitem{deltaG}
High-Field Magnetoresistance of Organic Semiconductors
G. Joshi, M. Y. Teferi, R. Miller, S. Jamali, M. Groesbeck, J. van Tol, R. McLaughlin, Z. V. Vardeny, J. M. Lupton, H. Malissa, and C. Boehme, "High-Field Magnetoresistance of Organic Semiconductors,"
Phys. Rev. Applied {\bf 10}, 024008 (2018).


\bibitem{Boehme+Lips}
C. Boehme and K. Lips,
"Theory of time-domain measurement of spin-dependent recombination with pulsed electrically detected magnetic resonance,"
Phys. Rev. B {\bf 68}, 245105  (2003).

\bibitem{Boehme+Lupton+Rabi}
D. R. McCamey, K. J. van Schooten, W. J. Baker, S.-Y. Lee, S.-Y. Paik, J. M. Lupton, and C. Boehme,
"Hyperfine-Field-Mediated Spin Beating in Electrostatically Bound Charge Carrier Pairs,"
Phys. Rev. Lett. {\bf 104}, 017601 (2010).

\bibitem{HighB_0}
W. J. Baker, D. R. McCamey, K. J. van Schooten, J. M. Lupton, and C. Boehme,
"Differentiation between polaron-pair and triplet-exciton polaron spin-dependent mechanisms in organic light-emitting diodes by coherent spin beating"
Phys. Rev. B {\bf 84}, 165205 (2011).


\bibitem{Saam}
M. E. Limes, J. Wang, W. J. Baker, S.-Y. Lee, B. Saam, and C. Boehme, "Numerical study of spin-dependent transition rates within pairs of dipolar and exchange coupled spins with 
$s=1/2$ during magnetic resonant excitation,"
Phys. Rev. B {\bf 87}, 165204 (2013).

\bibitem{Glenn} R. Glenn, W. J. Baker, C. Boehme, and M. E. Raikh, "Analytical description of spin-Rabi oscillation controlled electronic transitions rates between weakly coupled pairs of paramagnetic states with $S = 1/2$,"
Phys. Rev. B {\bf 87}, 155208 (2013).











\end{thebibliography}
\end{document}